# High-pressure electronic states in semiconductors studied by infrared spectroscopy: metallization and band gap tuning in Mg$_2$Si, InAs and InSb

Hidekazu Okamura[1]*, Haruna Okazaki[1], Katsunori Marugaku[1], Subin Lee[1], Tomoki Yoneda[1], Haruhiko Udono[2], Yoshihisa Mori[3], Mitsuhiko Maesato[4], Hiroshi Kitagawa[4], Yuka Ikemoto[5], and Taro Moriwaki[5]

[1]*Department of Applied Chemistry, Tokushima University, Tokushima 770-8506, Japan*

[2]*Department of Electrical and Electronic System Engineering, Ibaraki University, Hitachi 316-8511, Japan*

[3]*Department of Applied Science, Okayama University of Science, Okayama 700-0005, Japan*

[4]*Division of Chemistry, Graduate School of Science, Kyoto University, Kyoto 606-8502, Japan*

[5]*Japan Synchrotron Radiation Research Institute, Sayo 678-1297, Japan*

*E-mail: ho@tokushima-u.ac.jp

In this article, a brief introduction is first given on infrared studies of materials at high pressures using a diamond anvil cell. Then, our recent results of high-pressure infrared studies are described for Mg$_2$Si, InAs, and InSb. For Mg$_2$Si, pressure-induced metallization at pressures near 10 GPa were clearly demonstrated for both carrier-doped and undoped Mg$_2$Si by large increases of reflectivity. For InAs and InSb, their band gap ($E_g$) increased rapidly and almost linearly with pressure with linear coefficients of $dE_g/dP =$ 84.6 and 112 meV/GPa, respectively. Obtained values of $E_g$ versus lattice parameter at high pressures are compared with those for other III-V semiconductors at ambient pressure, giving unique insight into effects of physical and chemical pressures on $E_g$. Above the structural transition pressures of 7 and 3 GPa for InAs and InSb, respectively, they exhibit highly metallic characteristics accompanied by high reflectivity.

## 1. Introduction

Application of high pressures is a useful technique to explore novel materials properties that cannot be realized at ambient pressure.[1] With an external pressure, important material parameters such as the interatomic distance and electronic hybridization can be tuned without causing disorder in the crystal lattice. As a result, for example, various structural phase transitions and sometimes amorphizations have been observed at high pressures.[2] The electronic state of a material is also a sensitive function of pressure. Pressure-induced transitions between metallic and insulating (semiconducting) states and those to superconducting states have been observed for many materials.[3] Various kinds of pressure generating devices have been developed and used. Among them, diamond anvil cells (DACs)[1] have the advantage of being compact and easy to handle. With DAC, one can generate high pressures up to tens of GPa rather easily, and even up to a few hundred GPa though it is more challenging. In addition, since diamond is transparent over a wide spectral range, various optical techniques can be used with DAC including Raman, visible/UV, and infrared (IR) spectroscopies. The pressure within the sample space in a DAC can be conveniently measured by the ruby fluorescence method,[4] where small ruby pieces are loaded with the sample and the peak shifts of their fluorescence are utilized to measure the pressure.

IR spectroscopy has been used extensively to probe the low-energy excitations near the Fermi level ($E_\mathrm{F}$) of a material.[5] For example, the dynamics of free carriers (Drude response), various energy gaps (band gap, superconducting gap, charge density wave gap, etc), and exciton states have been studied by IR spectroscopy. By combining IR spectroscopy with a DAC, one can study these phenomena as a function of pressure, and hence as functions of various material parameters mentioned above.[6] There are other well-known spectroscopic techniques such as photoemission and tunneling spectroscopies to study electronic states near $E_\mathrm{F}$. They cannot be used, however, for high pressure studies since they cannot be performed on a sample confined in a pressure cell such as DAC. Namely, IR spectroscopy combined with a DAC provides a unique opportunity to study the electronic states near $E_\mathrm{F}$ at high pressures, although it is technically challenging as discussed below.

We have developed high-pressure IR spectroscopy technique, which has already been reported in detail.[6] The sample space in a DAC typically has a diameter of 400 and 200 μm for studies up to 10 and 20 GPa, respectively, and the sample sizes are typically 100-150 μm. In such a restricted condition, it is challenging to precisely perform infrared reflection or transmission measurements using a crystal sample. This is especially the case in the far-IR range, where diffraction effects become significant due to the long wavelength (10-100 μm, corresponding to 100-1000 cm$^{-1}$ wavenumber range and 0.012-0.12 eV photon energy range). To overcome this difficulty, in many of our high-pressure IR studies, synchrotron radiation (SR) has been utilized as a bright IR source.[7] (The brightness of IR SR has been utilized actively at many SR facilities worldwide, not only for high-pressure research but also for many other research fields.[7]) We have performed high-pressure IR studies on a wide variety of materials such as rare-earth monochalcogenides (YbS,[8] YbTe, TmTe), filled skutterudites (CeRu$_4$Sb$_{12}$,[9] PrRu$_4$P$_{12}$[10]), elemental Yb metal,[11] strongly correlated heavy-fermion metals (CeCoIn$_5$,[12] CeRhIn$_5$,[13] YbNi$_3$Ga$_9$[13], YbCu$_2$Ge$_2$[14]), an electronic ferroelectric (LuFe$_2$O$_4$),[15] pressure-induced superconductors (SrFe$_2$As$_2$,[16] AuTe$_2$[17]), an organic molecular insulator [β'-(BEDT-TTF)$_2$ICl$_2$],[18] an excitonic insulator (Ta$_2$NiSe$_5$),[19,20] and molecular clathrate hydrates.[21,22]

In this paper, we describe our recent results on high-pressure IR studies of semiconductors, namely Mg$_2$Si, InAs, and InSb. Mg$_2$Si has recently attracted much attention for thermoelectric[23] and optical[24] applications. It has also exhibited interesting high-pressure properties including an enhancement of thermoelectric power factor at high pressures.[25] On the other hand, InAs and InSb are narrow-gap semiconductors already widely used for practical devices such as IR detectors. In addition, their crystal structures at high pressures have attracted a lot of interest.[26] Rather surprisingly, however, their electronic states at high pressures have been much less studied experimentally. We have studied the infrared spectra of these compounds at high pressures up to 16 GPa. For Mg$_2$Si, our results indicate that the band gap is progressively reduced by pressure, and finally disappears at around 10 GPa. For InAs and InSb, in contrast, the band gap increases with pressure, but it suddenly disappears above the critical pressure where a structural phase transition occurs. These results are described in detail below.

## 2. Experimental methods

Figure 1 illustrates IR reflection and transmission studies at high pressures using a DAC. A pair of diamond anvils and a hole through a metal gasket are used to seal a sample with small ruby pieces, the latter of which are used to monitor the pressure as already explained above. Samples used were thin plates of single crystals with typical sizes of 150×150×35 μm$^3$ prepared by mechanical polishing. A flat and specular surface of a sample was in direct contact with the culet surface of a diamond anvil. The sample space was filled with a pressure transmitting medium, which was CsI for the far-IR range and NaCl for the other ranges. For a reflection study, a gold film was also placed in the DAC as a reference of reflectivity. Diamond anvils with culet diameters of 800 and 600 μm were used for studies up to 10 and 20 GPa, respectively. For the results reported in this paper, SR was used in the far-IR range at the beamline BL43IR of SPring-8.[27] At BL43IR, a custom-made IR microscope is available for both mid- and far-IR studies at high pressures with an *in-situ* pressure monitoring capability.[6] For mid-IR studies, high-pressure IR apparatus at Tokushima University was used. More details of our high-pressure IR techniques including data analysis methods[6,28] and more information about the IR-SR research[7] have already been described.

## 3. Results and discussion

### 3.1 Pressure-induced metallization in carrier-doped and undoped Mg$_2$Si

Mg$_2$Si is a semiconductor with a band gap of 0.61 eV,[24] which has attained increasing interest due to potential applications to infrared detectors[24] and thermoelectric devices.[23,24] Among the known thermoelectric materials, Mg$_2$Si has the advantage of being non-toxic, in contrast to several well-known thermoelectric materials containing toxic elements such as Pb and Te. In addition, both Mg and Si are abundant near the surface crust of the earth. To introduce charge carriers into Mg$_2$Si, both *n*-type and *p*-type dopings have been performed.[23] Interestingly, *n*-type Mg$_2$Si doped with 1% Al exhibited an enhancement of thermoelectric power factor under pressure with a maximum at about 2 GPa.[25] Subsequently, a high-pressure XRD and IR study of Al-doped Mg$_2$Si was performed.[29] In particular, the mid-IR reflectivity was observed to increase significantly with pressure above about 10 GPa.[29] Since these studies were performed on carrier-doped Mg$_2$Si, it was unclear whether the pressure response of the doped free carriers was important or that of intrinsic Mg$_2$Si itself was important. Therefore, we have made IR reflectivity studies of both doped and undoped Mg$_2$Si at high pressures. In addition, the present study has been made on bulk crystal samples, rather than on powder samples as in the previous study.[29] The 1% Al-doped Mg$_2$Si crystals were obtained from Union Materials Inc., while the undoped, high-purity Mg$_2$Si single crystals were grown by Bridgman method using a pyrolytic graphite crucible as previously described.[24] NaCl and CsI were used as pressure transmitting medium for mid-IR and far-IR studies, respectively.[30]

Figure 2(a) shows reflectivity spectra of 1% Al-doped Mg$_2$Si at high pressures and at room temperature.[31] At a low pressure of 0.5 GPa, the reflectivity increases sharply below an onset at about 0.13 eV (indicated by the blue arrow). This high reflection band is caused by a plasma oscillation of free carriers, and the onset of reflectivity corresponds to the plasma

frequency,[32)]

$$\omega_p = \left(\frac{ne^2}{\epsilon_\infty m^*}\right)^{1/2}, \qquad (1)$$

where $n$ is the carrier density, $e$ is the elementary charge, $\epsilon_\infty$ is the dielectric constant given by interband transitions, and $m^*$ is the effective mass. The reflectivity spectrum measured at ambient pressure (not shown) also had a plasma edge at 0.13 eV, which shows that the high reflection is caused by free electrons introduced by the Al doping. $\epsilon_\infty = 13.2\epsilon_0$ was obtained from analyzing the reflectivity spectrum at ambient pressure (not shown here) using the Drude-Lorentz oscillator model,[32)] and the electron effective mass of $m^* \approx 0.46 m_0$ has been reported.[33)] With these values, setting $\hbar\omega_p = 0.13$ eV in Eq. (1) yields $n = 7.5 \times 10^{19}$ cm$^{-3}$ for the electron density at ambient and low pressures. The sharp feature marked with the asterisk in Fig. 2(a) is due to optical phonons. With increasing pressure above 8 GPa, the reflectivity rapidly increases, which indicates that the carrier density becomes much larger. This result strongly suggests that the band gap is closed and Mg$_2$Si becomes metallic at around 10 GPa. The onset rapidly shifts to higher energy with further increasing pressure, as indicated by the vertical arrows in Fig. 2(a). The carrier density calculated from $\hbar\omega_p$ (onset energy) as discussed above is plotted versus pressure in Fig. 2(b). It is seen that the carrier density increases significantly after the band gap is closed, and reaches about $n = 1.4 \times 10^{21}$ cm$^{-3}$ which is about 20 times the ambient pressure value. While the observed shifts of reflectivity onset are qualitatively similar to those in the previous study,[29)] the reflectivity here is much higher probably because a bulk crystal sample was used in the present work.

Next, Fig. 3 displays reflectivity spectra of undoped Mg$_2$Si measured in the far-IR range at high pressures and at room temperature. Note that, since the sample is undoped, a plasma reflection band is not observed at low pressures in contrast to the case of Al doped Mg$_2$Si in Fig. 2(a). The two peaks located at 37 and 42 meV result from optical phonons. Besides the phonon peaks, the reflectivity is low and flat at 0.5 GPa due to the insulating electronic state of undoped Mg$_2$Si. With increasing pressure, however, the reflectivity increases gradually, and then quite rapidly at pressures between 8 and 10 GPa. The pressure range of this rapid increase almost overlaps with that for the Al-doped sample in Fig. 2(a). Namely, the increase of reflectivity due to metallization at around 10 GPa is an intrinsic property of Mg$_2$Si observed even without doped carriers. Our result for undoped Mg$_2$Si is very consistent with the previous electrical resistivity study at high pressures.[34)] Namely, the resistivity of undoped $Mg_2Si$ progressively decreased with pressure to 10 GPa, and it decreased much more rapidly at 10-13 GPa, suggesting an energy gap closing at 10 GPa.[4] Note also that the phonon peaks in Fig. 3, which are still clearly visible at 9 GPa, seem to be significantly broadened at 10 GPa. This result may suggest a change in the crystal symmetry, and hence a structural transition. There have been mixed experimental results on whether undoped $Mg_2Si$ undergoes a structural transition at pressures around 10 GPa[35-37] or not,[29] depending on the experimental conditions such as the types of pressure cell (DAC[29,35,36] or multi-anvil cell[37]) and pressure transmitting medium (silicone oil,[35] methanol-ethanol mixture,[36] NaCl,[36] or He[29]). Among them, a high-pressure study using DAC and NaCl,[36] which is also the case in our study, did observe a structural transition above 10.1 GPa. It is therefore likely that a structural transition actually occurred in our study at about 10 GPa. Theoretical band calculations for $Mg_2Si$ at high pressures have also been reported,[38,39] suggesting metallic states as actually observed here. To study the electronic states in undoped $Mg_2Si$ above 10 GPa in more detail, our work is in progress to extend the measurement to mid-IR range and to higher pressures.

3.2 Pressure tuning of the band gap and metallization in InAs and InSb

As already mentioned, InAs and InSb are well known narrow-gap semiconductors. At ambient condition, they both have the zinc blende crystal structure with lattice parameters $a_0 =$ 6.0583 and 6.4794 Å with their band gaps $E_g = $ 0.35 eV and 0.17 eV, respectively. Like many other III-V compounds, their crystal structures at high pressures have thoroughly been studied.[26] With increasing pressure, InAs undergoes a phase transition into a rock salt structure above $P_c \approx$ 7 GPa with a 17% volume reduction, and InSb to an orthorhombic structure above $P_c \approx$ 3 GPa.[26] In the high-pressure phase above $P_c$, InAs and InSb have been known to be metallic, but their electronic states have not been well studied experimentally. To probe the pressure evolution of their electronic states, we have studied their absorption and reflection spectra to 10 GPa using DAC. The thin-plate samples of InAs and InSb used with DAC had dimensions of approximately 150×150×30 μm³. They were obtained by mechanically polishing commercially available wafers of single crystal, which were not intentionally doped. The high-pressure IR studies with DAC were carried out at room temperature using NaCl as the pressure transmitting medium.

Figure 4(a) shows the absorption spectra of InAs at high pressures. At 0.1 GPa, the absorbance rapidly increases with an onset at 0.35 eV. Since InAs is a direct gap semiconductor, the onset energy should correspond to the band gap, $E_g$. (Note that exciton-related spectral structures are not observed here.) With increasing pressure, the onset shifts to higher energy, which indicates that $E_g$ in InAs increases with pressure, in contrast to the case of Mg$_2$Si discussed above. The onset keeps shifting with pressure up to 7.1 GPa, but at 7.5 GPa, an onset is absent and the absorbance is high at all the measured photon energies. Since the pressure where the onset disappears almost coincides with $P_c$, our results indicate that the band gap disappears at $P_c$, and that the high-pressure phase above $P_c$ is metallic, which is also demonstrated by the reflectivity results discussed later. To obtain the value of $E_g$, we assume that $E_g$ is given by the onset of absorbance, which is estimated as illustrated by the broken lines in Fig. 4(a). Similar estimates are also made for data at other pressures, and in Fig. 4(b) the obtained values of $E_g$ are plotted as a function of pressure ($P$). It is seen that $E_g$ increases almost linearly with $P$, up to the maximum $E_g$ of 0.86 eV. Results of two sets of data are very similar to each other below 5 GPa with a linear pressure coefficient of $\alpha_P = dE_g/dP =$ 84.6 meV/GPa. The pressure dependences of $E_g$ and $\alpha_p$ for InAs have been considered by many studies.[40-44] Experimentally, an optical absorption study reported $\alpha_p =$ 48 and 32 meV/GPa below 2 GPa and at 2-5 GPa, respectively,[40] which are much smaller than the present result. Another experimental study based on electrical resistivity and Hall effect measurements reported $\alpha_p =$ 114 meV/GPa below 1.5 GPa and at room temperature.[41] The early study[40] is pioneering but quite old, and the transport study[41] is done over a wide temperature range and is based on many assumptions about material parameters to determine $\alpha_p$. In contrast, our method is quite simple and straightforward, and we believe that our result is reliable. Theoretically, $\alpha_p =$ 95.6[42], 103.5[43], and 110[44] meV/GPa have been reported by theoretical band calculations. Note that all these theoretical results are larger than the present experimental one of 84.6 meV/GPa, with 15-30% differences.

Figures 5(a) and 5(b) display results on InSb. The absorption onset shifts to higher energy with pressure up to 2.8 GPa, but is absent at 3.0 GPa. $E_g$ estimated from the absorbance increases almost linearly with pressure from 0.17 eV at ambient pressure to 0.51 eV at 2.8 GPa. Two independent sets of data exhibit slightly different $dE_g/dP$, but they are very close and their average yields a pressure coefficient of $\alpha_P =$ 112 meV/GPa, which is about 30% larger than that for InAs. Regarding previous studies of $\alpha_p$ for InSb, to the best of our knowledge, no experimental study has been reported so far,[45,46] and the present result of $\alpha_p =$ 112 meV/GPa seems to be the first experimental report. Theoretically, $\alpha_p$ values of 136.7[42] and 132.7[47] meV/GPa have been reported using first-principles band structure calculations. These theoretical values are about 20% larger than the present experimental value of 112 meV/GPa, similarly to the case of InAs above. The reason why the theoretical studies systematically overestimate $\alpha_p$ is unclear at present. Further theoretical studies seem necessary to fully understand the difference between experimental and theoretical results.

To compare the results between InAs and InSb, in Fig. 6(a) we plot $E_g$ versus $P$ together, and in Fig.6(b) we plot the normalized band gap, $E_g(P)/E_g(0)$, versus the normalized pressure, $P/P_c$. Here, $P_c$ was set to 7 and 3 GPa for InAs and InSb, respectively. It is evident that InAs and InSb exhibit almost the same $P$ dependence of $E_g$ when they are normalized this way. This result mainly results from the fact that the maximum $E_g$ before reaching $P_c$ is about 2.7 times $E_g(0)$ for both InAs and InSb. From Fig. 6(b), it appears as if $E_g$ is an order parameter in the semiconductor-to-metal phase transitions of InAs and InSb. Since they share the same crystal structure with similar electron configurations, the above result likely results from a common, universal mechanism of band gap response against compression.

Next, we compare the evolution of $E_g$ with the lattice contraction, $-\Delta a/a_0$, rather than with $P$. To do so, we utilize the Murnaghan equation of state,[48]

$$P = (B_0/B_0')\left[(a/a_0)^{-3B_0'} - 1\right], \qquad (2)$$

where $a$ and $a_0$ are the lattice parameter and its ambient-pressure value, $B_0$ and $B_0'$ are the bulk modulus and its pressure derivative, respectively. To obtain $-\Delta a/a_0$ values at different pressures for InAs and InSb,[49] we used Eq. (2) with the following values of $B_0$ and $B_0'$: 57.9 GPa and 4.79 for InAs,[50] and 45.7 GPa and 4.89 for InSb,[51] respectively. Figs. 7(a) and 7(b) plot $E_g$ and normalized $E_g$ versus $-\Delta a/a_0$, respectively. In Fig. 7(a), both InAs and InSb have almost the same slope versus $-\Delta a/a_0$. This is because InSb has larger pressure coefficient of $E_g$ but is more compressible with a smaller $B_0$ than InAs. In Fig. 7(b), InSb has a larger slope of normalized $E_g$ versus $-\Delta a/a_0$, which should be a direct consequence of larger compressibility for InSb.

We now compare the effects of pressure on $E_g$ observed for InAs and InSb with those of element substitution. This corresponds to a comparison between effects of physical and chemical pressures on $E_g$. In Fig. 8, we plot published $E_g$ values of various III-V binary compounds and their ternary alloys versus $a$ at ambient condition,[52] together with the high-pressure data of InAs and InSb obtained here. For InAs, its pressure dependence is located near and between the In → Ga and As → P substitution curves. This result seems quite reasonable because in the groups III and V elements, Ga and P are next to In and As with a smaller atomic radius, respectively. Namely, for InAs, pressure has similar effects on $E_g$ with a partial substitution by the next smaller element in the same group. This similarity suggests that the variation of $E_g$ with composition in (In, Ga)As and In(As, P) alloys should be mainly characterized by the lattice contraction. Regarding InSb, the pressure dependence is located near and above the In → Ga curve, which is similar to the case of InAs. However, the pressure dependence is located far above the Sb → As curve. The In(As, Sb) curve in Fig. 8 has a minimum with a large bowing parameter. This property has been well known, and electronic structures and $E_g$ in In(As, Ga) alloys have been studied in detail.[53] Although $a$ of In(As, Sb) varies almost linearly with composition, $E_g$ does not vary linearly with composition and exhibits a minimum as seen in Fig. 8. This variation of $E_g$ is strongly affected by reduction of spin-orbit band splitting with Sb → As replacement, which is due to much larger spin-orbit coupling of Sb than that of As.[53] Note that the Ga(As, Sb) curve also has a similar shape with a minimum due to similar reasons.

To probe the electronic states of InAs and InSb at $P > P_c$, we have measured their reflectivity spectra at high pressures as displayed in Figs. 9(a) and 9(b), respectively. It is seen that, with increasing $P$ above $P_c$, the reflectivity rapidly increases within a pressure range of about 1 GPa for both compounds. The high reflectivity apparently result from plasma reflection by free carriers, and demonstrate that these compounds in the high-pressure phase are metallic. Note that, unlike the case of $Mg_2Si$ in Fig. 2, the high reflectivity is observed over the entire infrared range, and the onset of reflectivity (plasma edge) is located above the IR photon energy range. Hence, we further performed high-pressure reflectivity studies in the visible-UV range up to 3.5 eV. A visible-UV spectromicroscope (JASCO MSV-370) was used with a DAC. A silver film was placed in the DAC as a reference. Fig. 10 shows the results for InAs.[54] At pressures below $P_c$, the reflectivity is low, and double peaks are observed at 2.5-3 eV range. These peaks, which are due to direct interband transitions at high energies, shift to higher energy with pressure as reported previously.[55] Near and above $P_c$, however, the reflectivity increases significantly with pressure, not only in the near-IR and visible ranges but also in the UV range. We identify the deep minimum of reflectivity at about 1.7 eV as the onset of reflectivity (plasma edge). To estimate the carrier density from the onset energy, we need to know $m^*$ of the carriers in the high-pressure phase with rock salt structure. A band calculation study of InAs with the rock salt structure has been reported,[56] but it does not provide information about $m^*$

Hence, we calculate the effective carrier density $n^* = n/(m^*/m_0)$, instead of $n$ itself, from Eq. (1). Also, the ambient pressure value of $\epsilon_\infty = 11.5\,\epsilon_0$ is used.[57] Then, Eq. (1) and $\hbar\omega_P = 1.7$ eV yield an effective carrier density of $n^* = 2.4 \times 10^{22}$ cm$^{-3}$. This is smaller but comparable with that of good metals such as gold, which has a free electron density of $n = 5.8 \times 10^{22}$ cm$^{-3}$ assuming one free electron per atom. Hence, our result suggests a highly metallic character of the high-pressure phase in InAs.[58]

Very recently, detailed transport studies of InAs at high pressures have been reported.[59] They report a reduction of electrical resistivity by four orders of magnitude with increasing pressure through $P_c$. They also report a carrier density of about $n = 3 \times 10^{18}$ cm$^{-3}$ above $P_c$ (10 GPa) from Hall effect. This density is much lower than that obtained in the present work, and in fact seems too low to account for the high reflectivity over a wide photon energy range observed at $P > P_c$. However, this is not necessarily a contradiction since the carrier density given by Hall effect is the difference between majority and minority carriers while that given by optical plasma frequency is the total of electrons and holes.

We briefly mention hystereses observed during our high-pressure optical studies on InAs and InSb. In Fig. 9, the reflectivity spectra recorded during decompression are also indicated (pink curves). They are very similar to those recorded at low pressures before compression to above $P_c$. Namely, the initial reflectivity spectrum before compression to above $P_c$ is recovered even after a decompression from above $P_c$ to a near-ambient pressure. However, a significant hysteresis was observed between compression and decompression: For InAs (InSb), a rapid increase of reflectivity during compression was observed at 7-8 GPa (3-4 GPa), while a large decrease during decompression was observed at 1-3 GPa (0.5-1.5 GPa). For absorption spectra, however, the transmission at photon energies below $E_g$ before compression to above $P_c$ was not recovered after decompression from above $P_c$ to ambient pressure. (Samples after decompression from above $P_c$ did not exhibit a transmission even at photon energies below $E_g$ at ambient pressure, in contrast to the case before compression.) The reason for these results are unclear and deserves further studies.

## 4. Conclusions

We have described our optical studies of semiconductors Mg$_2$Si, InAs, and InSb at high pressures, where a DAC has been used to generate pressure and an IR synchrotron radiation has been used when necessary. Variations of absorption and reflectivity spectra with pressure have been presented and discussed in terms of the evolution of electronic states in these compounds.  In particular, pressure-induced semiconductor-to-metal transitions and variations of the band gap have been discussed. For the reflectivity data presented in this work, to analyze the pressure evolution of electronic states more quantitatively, they should be further analyzed with the Kramers-Kronig relations to derive and evaluate the optical functions such as dielectric function and optical conductivity, as previously done for other compounds.[8-20] Such work is in progress and should be presented elsewhere.

## Acknowledgments

This study was supported by the Japan Society for the Promotion of Science (JSPS) KAKENHI Grant Number 24K06962.

with DAC was about 20 meV (160 cm$^{-1}$). Therefore, CsI was chosen for the far-IR study to avoid reflection signal from pressure transmitting medium. (Note that the phonon peaks also shift to the higher-energy side with pressure.)

**Figures**

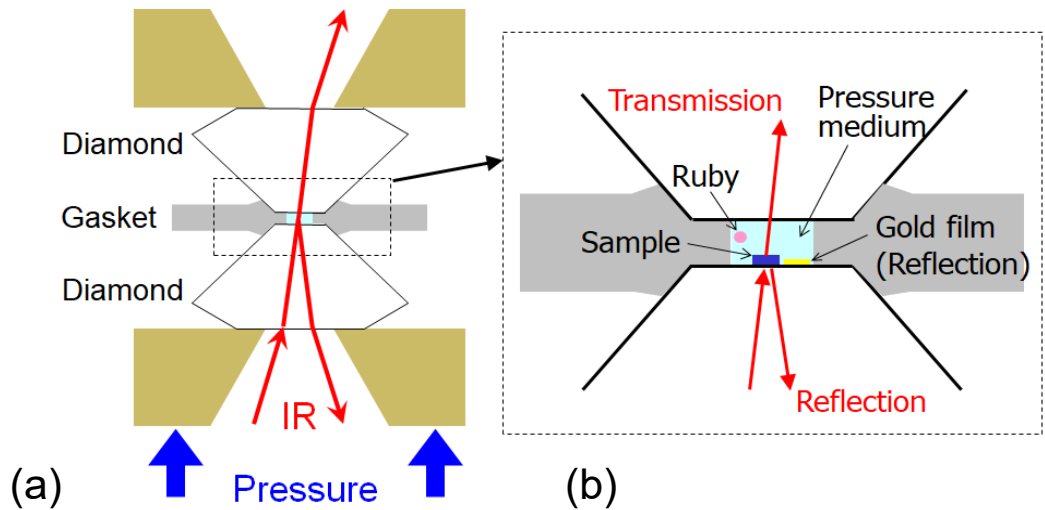

**Fig. 1.** Schematic description of IR reflection and transmission studies at high pressures using a DAC. Note that, in (b), a gold film is used as a reference of reflectivity. For a transmission study, the gold film is not placed there, and the IR transmission without going through the sample (going through the pressure transmitting medium only ) is used as a reference.

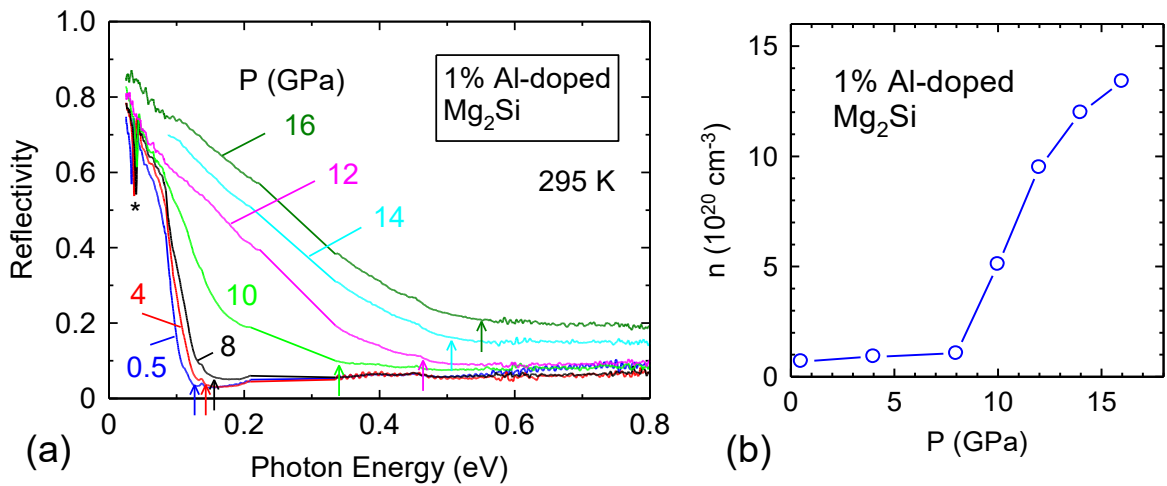

**Fig. 2.** (a) Infrared reflectivity spectra of 1% Al-doped $Mg_2Si$ at high pressures and at room temperature.[31] The "*" mark indicates structures due to optical phonon, and the vertical arrows indicate the approximate positions of the onset of high reflectivity (plasma reflection) due to free carriers. Note that the spectra are interpolated at 0.2-0.34 eV range since the spectra could not be measured well due to strong absorption by diamond. (b) Pressure dependence of the carrier density ($n$) versus pressure, which is calculated from the data in (a) as described in the text.

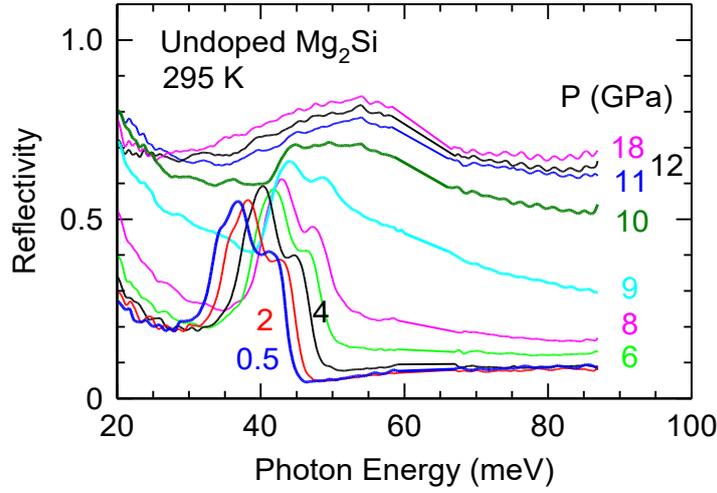

**FIG. 3.** Reflectivity spectra of undoped Mg$_2$Si in the far-IR range at high pressures and at room temperature.

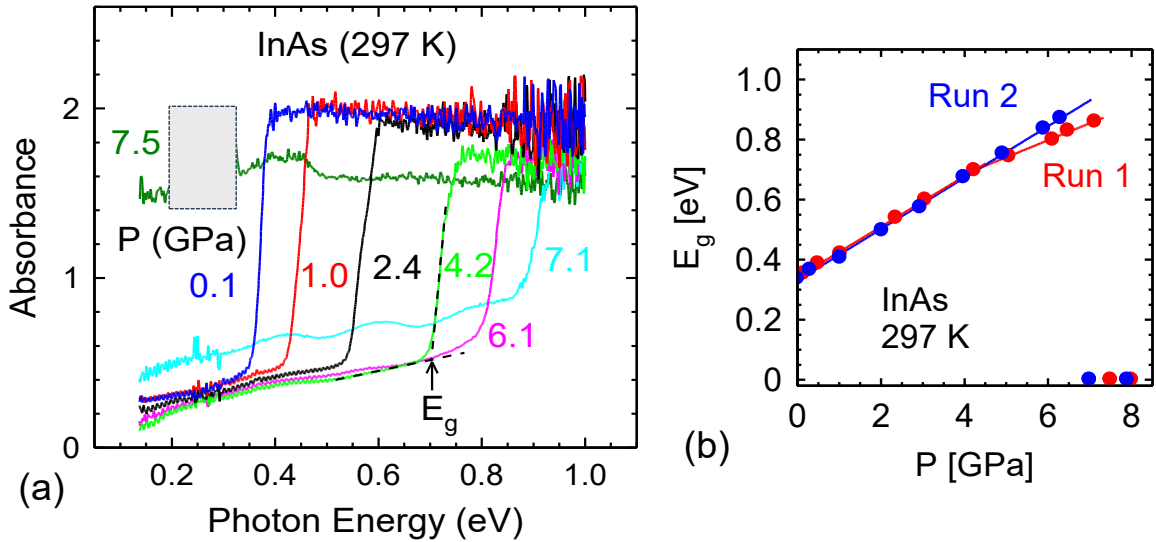

**FIG. 4.** (a) Absorption spectra of InAs at high pressures ($P$). The spectral range indicated by the gray box could not be measured well due to strong absorption by diamond. The broken lines illustrate how the band gap ($E_g$) is estimated for the 4.2 GPa data. (b) Obtained values of $E_g$ plotted versus $P$. The pressure coefficient is almost the same for the two sets of data at pressures below 5 GPa and is obtained as 84.8 meV/GPa.

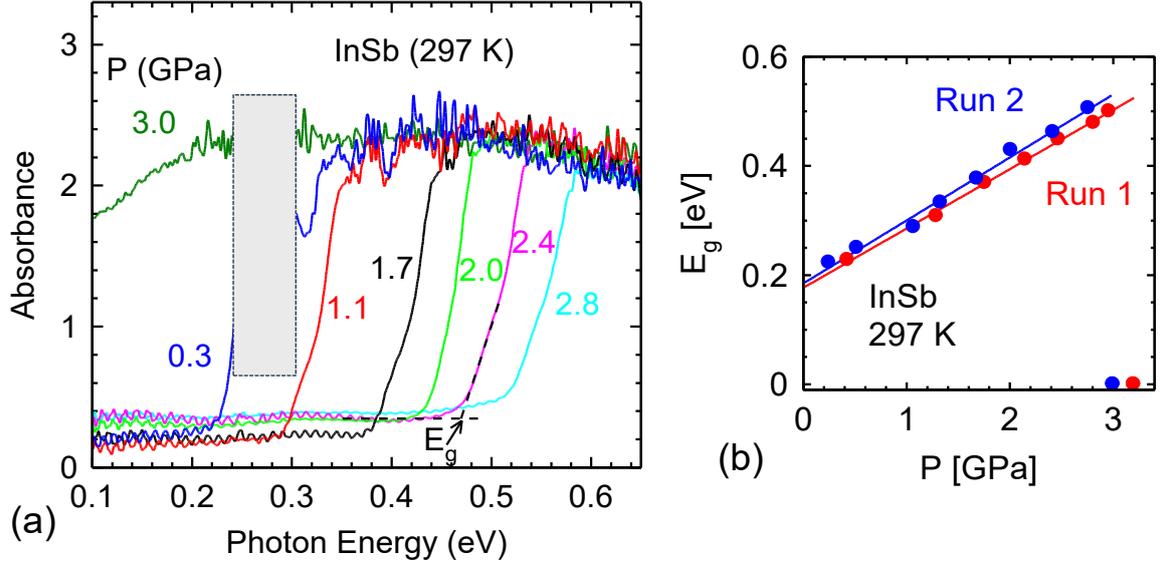

**FIG. 5.** (a) Absorption spectra of InSb at high pressures ($P$). The spectral range indicated by the gray box could not be measured well due to strong absorption by diamond. The broken lines illustrate how the band gap ($E_g$) is estimated for the 2.4 GPa data. (b) Obtained values of $E_g$ plotted versus $P$. The $P$ coefficient of $E_g$ averaged over the two sets of data is obtained as 112 meV/GPa.

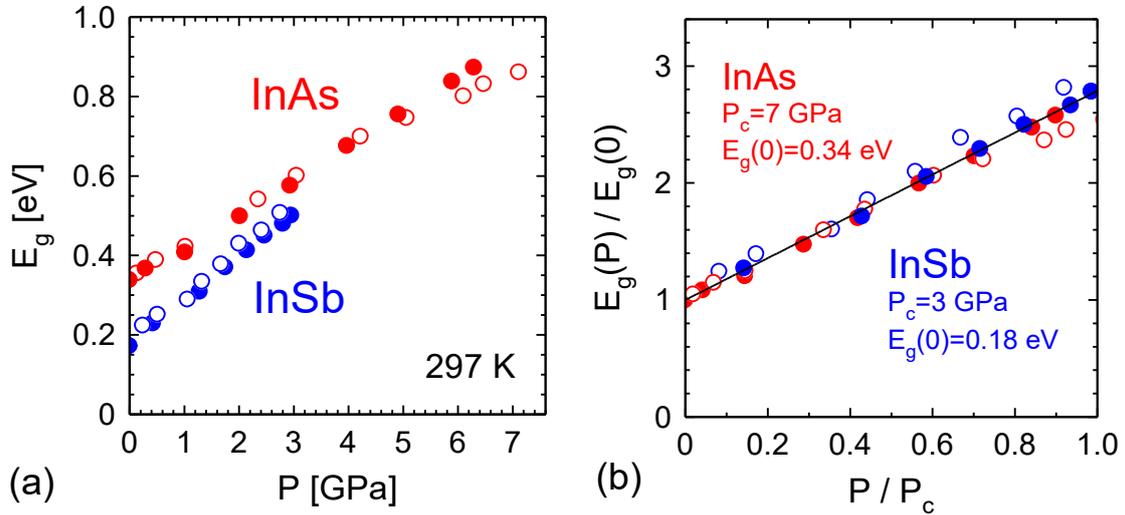

**FIG. 6.** (a) Comparison of the pressure dependence of $E_g$ between InAs and InSb. Filled and empty circles correspond to the same two sets of data in Figs. 4 and 5 for each compound. (b) Normalized band gap $E_g(P)/E_g(0)$ plotted versus the normalized pressure ($P/P_c$). The straight line is guide to the eye.

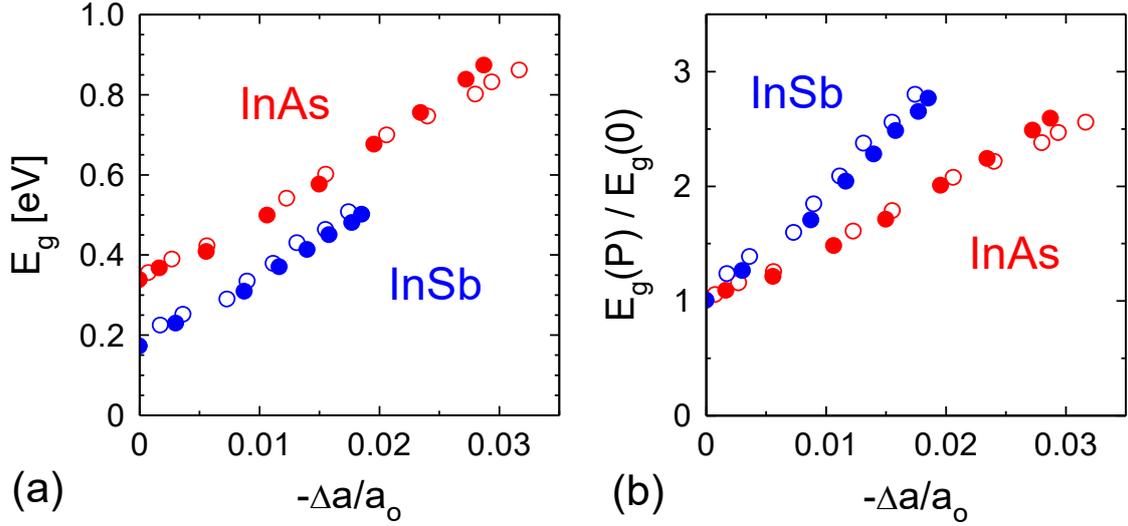

**FIG. 7.** Measured $E_g$ (a) and $E_g(P)/E_g(0)$ (b) in InAs and InSb plotted versus the lattice contraction $(-\Delta a/a_0)$.

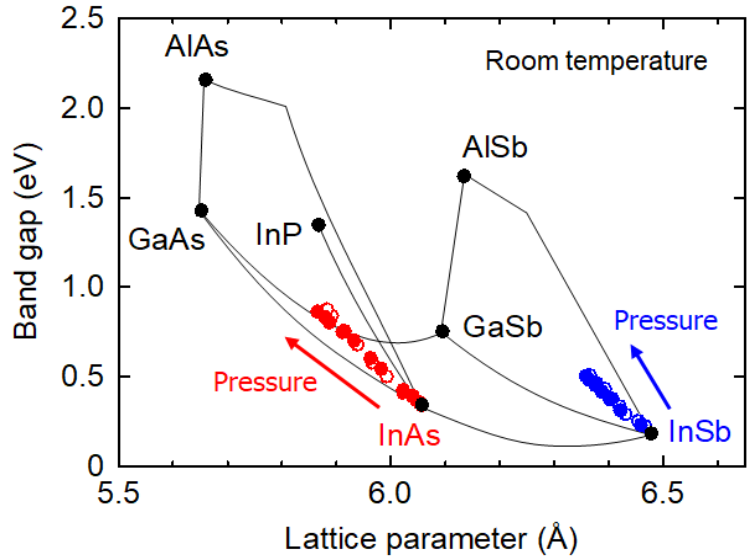

**FIG. 8.** Band gap versus lattice parameter for some III-V compounds and their alloys with the zinc blende structure at ambient condition, taken from a similar diagram in Ref. 52, compared with those of InAs and InSb at high pressures obtained in this work.

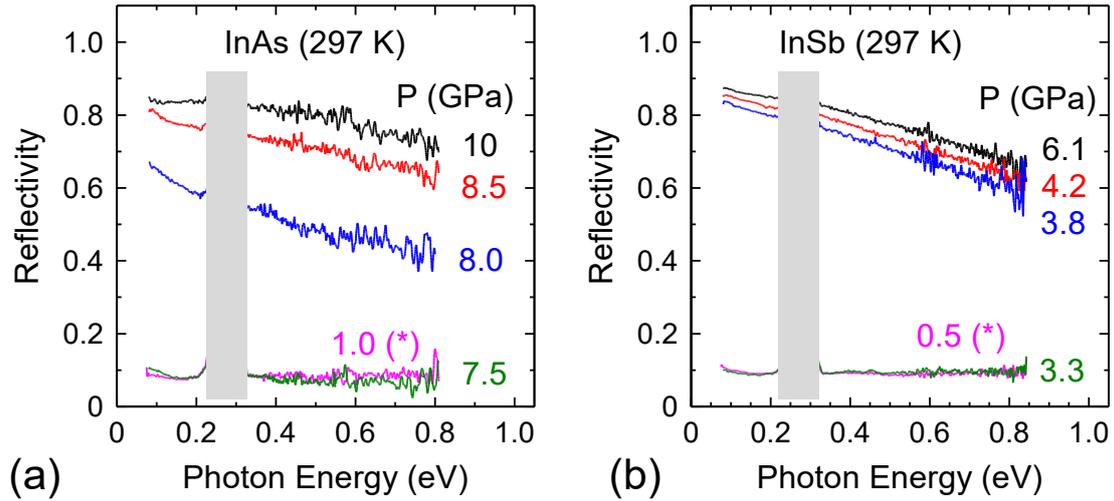

**FIG. 9.** Reflectivity spectra of (a) InAs and (b) InSb in the IR range measured at high pressures. The spectra whose pressures are marked with (*) were measured during decompression, while the others were measured during compression. Spectra at many other pressures were also recorded, but for clarity, only a small number of selected spectra are displayed here. The gray shaded spectral ranges could not be measured well due to strong absorption by diamond.

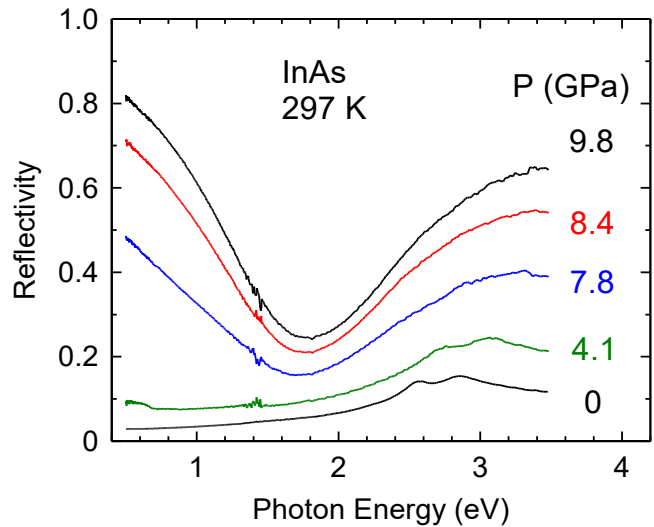

**FIG. 10.** Reflectivity spectra of InAs in the near-IR, visible, and UV ranges measured at high pressures.[54]